\documentclass[aps,preprint,groupedaddress,amsmath,amssymb,longbibliography]{revtex4-1}
\usepackage{graphicx}
\usepackage{color}
\usepackage{bm}% bold math
\usepackage{hyperref}
\usepackage{soul}
\usepackage{lineno}
\DeclareMathOperator{\tr}{Tr}

\begin{document}

\title{Interplay of Structure, Elasticity and Dynamics in Actin-Based Nematic Materials}

% Use letters for affiliations, numbers to show equal authorship (if applicable) and to indicate the corresponding author
\author{Rui Zhang$^{1}$}
\thanks{Equal contribution}
\author{Nitin Kumar$^{2,3}$}
\thanks{Equal contribution}
\author{Jennifer Ross$^{4}$}
\author{Margaret L. Gardel$^{2,3,5}$}
\email{gardel@uchicago.edu}
\author{Juan J. de Pablo$^{1,6}$}
\email{depablo@uchicago.edu}
\affiliation{
  $^1$Institute for Molecular Engineering, The University of Chicago, Chicago, Illinois 60637, USA \\
  $^2$James Franck Institute, The University of Chicago, Chicago, Illinois 60637, USA \\
  $^3$Department of Physics, The University of Chicago, Chicago, Illinois 60637, USA \\
  $^4$Department of Physics, University of Massachusetts Amherst, Amherst, Massachusetts 01003, USA \\
  $^5$Institute for Biophysical Dynamics, The University of Chicago, Chicago, Illinois 60637, USA \\
  $^6$Institute for Molecular Engineering, Argonne National Laboratory, Lemont, Illinois 60439, USA
  }
  
  \date{\today}

\begin{abstract}
Achieving control and tunability of lyotropic materials has been a long-standing goal of liquid crystal research. Here we show that the elasticity of a liquid crystal system consisting of a dense suspension of semiflexible biopolymers can be manipulated over a relatively wide range of elastic moduli. Specifically, thin films of actin filaments are assembled at an oil-water interface. At sufficiently high concentrations, one observes the formation of a nematic phase riddled with $\pm1/2$ topological defects, characteristic of a two-dimensional nematic system. As the average filament length increases, the defect morphology transitions from a U-shape into a V-shape, indicating the relative increase of the material's bend over splay modulus. Furthermore, through the sparse addition of rigid microtubule filaments, one can further control the liquid crystal elasticity. We show how the material's bend constant can be raised linearly as a function of microtubule filament density, and present a simple means to extract absolute values of the elastic moduli from purely optical observations. Finally, we demonstrate that it is possible to predict not only the static structure of the material, including its topological defects, but also the evolution of the system into dynamically arrested states. Despite the non-equilibrium nature of the material, our continuum model, which couples structure and hydrodynamics, is able to capture the annihilation and movement of defects over long time scales. Thus, we have experimentally realized a lyotropic liquid crystal system that can be truly engineered, with tunable mechanical properties, and a theoretical framework to capture its structure, mechanics, and dynamics.
\end{abstract}

% Optional adjustment to line up main text (after abstract) of first page with line numbers, when using both lineno and twocolumn options.
% You should only change this length when you've finalised the article contents.

\maketitle
\section{Introduction}
% If your first paragraph (i.e. with the \dropcap) contains a list environment (quote, quotation, theorem, definition, enumerate, itemize...), the line after the list may have some extra indentation. If this is the case, add \parshape=0 to the end of the list environment.
Recent advances have extended applications of nematic Liquid Crystals (LCs) onto realms that go beyond display technologies\cite{chen2011liquid} and elastomers\cite{warner2003liquid,camacho2004fast} to colloidal/molecular self-assembly\cite{zumer_science2006,amin_pnas,wang_natmat2015}, pathogen sensing\cite{lin_science2011,sadati_afm2015}, photonic devices\cite{humar2009electrically}, drug delivery\cite{zhang2016controlled}, and microfluidics\cite{sengupta2014liquid}. In LCs, topological defects correspond to locally disordered regions where the orientation of the mesogens\cite{deGennes_book} (the units that form an LC mesophase) changes abruptly. Defects exhibit unique optical and other physicochemical characteristics, and serve as the basis for many of the applications of these materials.
The microstructure of an LC, the so-called ``director field'', is determined by a delicate interplay between  elastic forces, geometrical constraints, and the influence of applied external fields\cite{tomar2012morphological,miller2013influence}. That microstructure dictates optical and mechanical response of the LC to external cues. Thus, the ability to control and optimize the LC elasticity is essential for emerging applications. The most widely used low-molecular-weight LCs are thermotropic (LC phases emerge at a certain temperature), and exhibit a nematic phase within certain, relatively narrow temperature range. In such LCs, the defect size is small ($\approx 10$~nm)\cite{rahimi_natcomm2017}, thereby placing severe constraints on applications that might rely on defects to perform certain functions (e.g. absorption and transport of colloidal particles\cite{amin_pnas}). Furthermore, in thermotropic systems, there is a limited range of accessible microstructures that are only achieved with significant changes in temperature\cite{deGennes_book}.

Lyotropic materials in which LC phases emerge at certain solute concentrations circumvent some of these shortcomings. In addition, certain lyotropic LCs are not toxic to many microbial species\cite{woolverton_liqcry2005} and neutral to antibody-antigen binding\cite{luk_chemmat2005}, providing possibility in biological applications\cite{park2012lyotropic}. Here we rely on suspensions of biopolymers to form lyotropic LCs, where long and semiflexible mesogens lead to the formation of nematic phases at sufficiently high concentrations. More specifically, we use filamentous actin (F-actin), with width of $\approx$ 7 nm and persistence length of 10-17 $\mu m$\cite{blanchoin2014actin,ott_pre1993}.
F-actin suspensions exhibit a nematic phase when the concentration rises above the Onsager limit\cite{viamontes2006isotropic,weirich2017liquid}. Here we deplete F-actin onto an oil-water interface, thereby forming a quasi-two-dimensional (2D) nematic system\cite{dogic_nature2012}. F-actin's average contour length can be tuned from $< 1 ~\mu m$ to $> 10 ~\mu m$ by addition of variable concentrations of a capping protein that limits polymer growth\cite{weeds_1993}. Because the corresponding topological defects that arise in nematic actin are on the order of $\mu m$, they can be visualized by optical microscopy. We demonstrate an approach to extract precise measurements of elastic moduli by examining the morphology of the $+1/2$ defects\cite{hudson1989frank}. We find that as the filament length increases from 1~$\mu $m to 2~$\mu $m, the defect morphology transitions from a U-shape to a V-shape, indicating that the nematic LC changes from a chromonic-LC-like (splay-dominant) material into a calamitic-LC-like (bend-dominant) system. We also demonstrate that the formation of a composite LC by addition of sparse concentrations of a more rigid mesogen leads to an increase of the bend elasticity of the resulting composite LC that grows linearly with its number density and average length.

Our 2D nematic system provides an ideal platform for detailed studies of the dynamics of defects in lyotropic LCs. We envisage applications in which topological defects could be used to transport cargo in a deterministic manner from one region to another. In order to engineer such systems, it is essential that a quantitative formalism be advanced with which to predict the dynamics of lyotropic LCs. Here we combine a Landau-de Gennes free energy model and the underlying momentum conservation equations and show that such an approach is capable of describing non-equilibrium dynamic processes in polydisperse, semiflexible biopolymer systems, paving the way for development of active systems that rely on defect motion to engender function.

\subsection*{Results and Discussion}
 \begin{figure*}[t]
 \centerline{\includegraphics[width=0.85\textwidth]{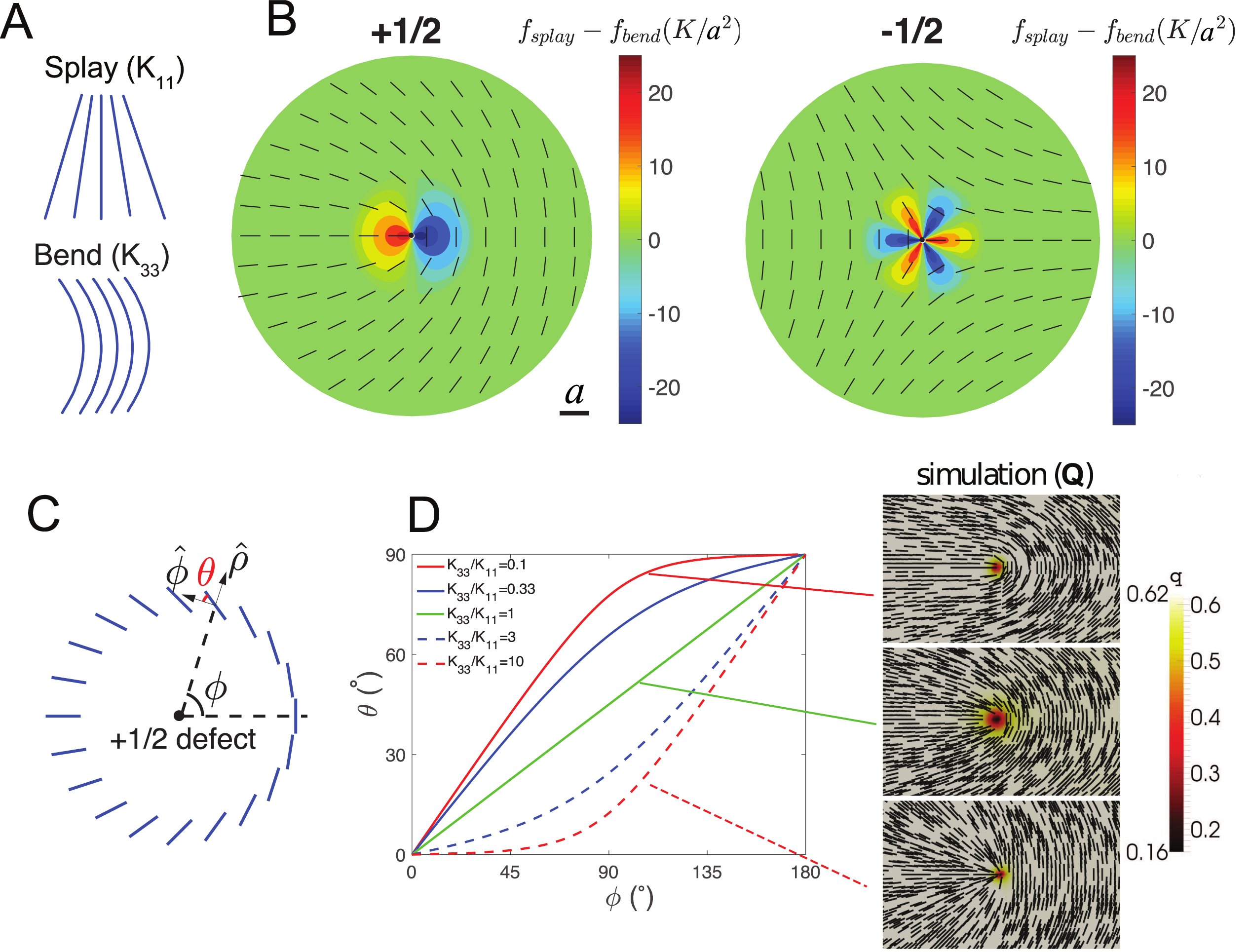}}
%\centering
%\includegraphics[width=0.85\linewidth]{fig1.eps}
\caption{\textbf{Morphology of $+1/2$ defects}. {\bf A}. Illustration of splay and bend distortion in nematic LC. {\bf B}. The director fields of $+1/2$ and $-1/2$ defect under one-elastic-constant approximation; The color indicates the difference in splay and bend energy density. $a$ is the unit length scale. {\bf C}. Quantitative description of defect morphology: $\phi$ is the polar coordinate; $\theta$ is the angle between the director and the angular vector ${\hat \phi}$. {\bf D}. The morphology of $+1/2$ defects as function of elastic constant ratio $\kappa\equiv K_{33}/K_{11}$. Images on the right side are the {\bf Q}-tensor based simulation results.}
\label{fig1}
\end{figure*}
The local average orientation of the mesogens is denoted by {\bf n}, which is a position dependent (unit) vector field that obeys nematic (head-tail) symmetry, i.e. {\bf n} and $-${\bf n} are undistinguishable. When an LC transitions from an isotropic phase to a nematic phase, topological defects emerge and annihilate over time as the material undergoes thermal equilibration. Topological defects appear as isotropic regions where the director field changes abruptly. Those defects can be categorized by their winding number or topological charge\cite{lavrentovich_book}. Fig.~\ref{fig1}B provides an illustration of $\pm1/2$ defects. The long-range order of nematic LCs is often characterized by the Frank-Oseen elastic theory\cite{frank1958liquid}, where the elastic energy density is given by
\begin{equation}\label{frank}
\begin{aligned}
f_e= \frac{1}{2}K_{11} &(\nabla\cdot {\bf n})^2+\frac{1}{2}K_{22}({\bf n}\cdot \nabla \times{\bf n})^2+\frac{1}{2}K_{33}({\bf n}\times (\nabla \times{\bf n}))^2 \\
&-\frac{1}{2}K_{24}\nabla\cdot [{\bf n}(\nabla\cdot{\bf n})+{\bf n}\times(\nabla\times{\bf n})].
\end{aligned}
\end{equation}
Here the four pre-factors penalize splay ($K_{11}$), twist ($K_{22}$), bend ($K_{33}$), and saddle-splay ($K_{24}$) deformations, respectively. A widely used simplification of the above expression is provided by the so-called ``one-constant-approximation'', which assumes $K_{11}=K_{22}=K_{33}=K_{24}\equiv K$. In that case, Eq.~\ref{frank} reduces to (using Einstein notation summing all repeating indices) $f_e=\frac{1}{2}K\partial_i n_j\partial_i n_j$, and captures many qualitative physics of nematic LCs.
In a 2D nematic system, mesogens are constrained to lie in a plane and the twist ($K_{22}$) and saddle-splay ($K_{24}$) deformation modes are irrelevant. Therefore the splay ($K_{11}$) and bend ($K_{33}$) influence the structure of topological defects, depicted in Fig.~\ref{fig1}A. By minimizing the system's total elastic energy over an area $S$ ($\int_S f_e dS$ with respect to ${\bf n}$), an analytical expression for the director field near a defect is obtained. Introducing an elastic distortion energy density according to: $f_{splay}=\frac{1}{2}K_{11}(\nabla\cdot{\bf n})^2$ and $f_{bend}=\frac{1}{2}K_{33}({\bf n}\times ( \nabla\times{\bf n}))^2$, we quantify the distribution of splay and bend distortions near a defect. In Fig.~\ref{fig1}B, $f_{splay}-f_{bend}$ near $\pm1/2$ defects is plotted under the one-constant-approximation, demonstrating that the $+1/2$ defect has two-fold symmetry, whereas the $-1/2$ defect has three-fold symmetry.

To better account for nematic symmetry, one can also construct a second order, symmetric, traceless tensor {\bf Q} based on {\bf n}. The {\bf Q}-tensor (under a uniaxial assumption) is defined by ${\bf Q}=q({\bf n n}-{\bf I}/3)$, where $q$ is the scalar order parameter that characterizes how well aligned the mesogens are. The nematic symmetry is automatically satisfied when a {\bf Q} tensor is introduced. Eq.~\ref{frank} can be written in terms of the {\bf Q} tensor and incorporated into the Landau-de Gennes free energy model, and a Ginzburg-Landau approach is implemented to arrive at the LC's equilibrium structure (see Methods).
To explore the fine structure of the topological defect, we go beyond the one-constant approximation and consider the bend-to-splay ratio $\kappa\equiv K_{33}/K_{11}$. We find that the analytical expressions derived are consistent with the results from full {\bf Q}-tensor simulations (see Supplementary Information).
To quantify the director field near a $+1/2$ defect, one can introduce an angle $\theta$ (illustrated in Fig.~\ref{fig1}C) in a polar coordinate system $(\rho,\phi)$, such that the director can be expressed as ${\bf n}=\hat{\bf \rho}\sin\theta+\hat{\bf \phi}\cos\theta$. In Fig.~\ref{fig1}D, $\theta(\phi)$ is plotted for various values of $\kappa$. These calculations show that the $+1/2$ defect morphology adopts a U-shape when $K_{33}<K_{11}$ ($\kappa<1$) and a V-shape when $K_{33}>K_{11}$ ($\kappa>1$).
When $\kappa <1$, splay distortion is more energetically demanding, whereas bend distortion is easier. The system therefore tries to squeeze any high-splay regions and expand high-bend regions. The resulting directors are more horizontal, leading to a U-shape morphology. In contrast, if $\kappa >1$, splay distortions become energetically easier and bend distortions become more demanding; the system is therefore stretched in the opposite direction. It tries to spread high-splay regions and suppress high-bend regions, resulting in a V-shape director field. One can also examine the core shape of a $-1/2$ defect by varying $\kappa$ (see Fig. \ref{s1} in the Supplementary Information).

 \begin{figure*}%[tbhp]
  \centerline{\includegraphics[width=0.85\textwidth]{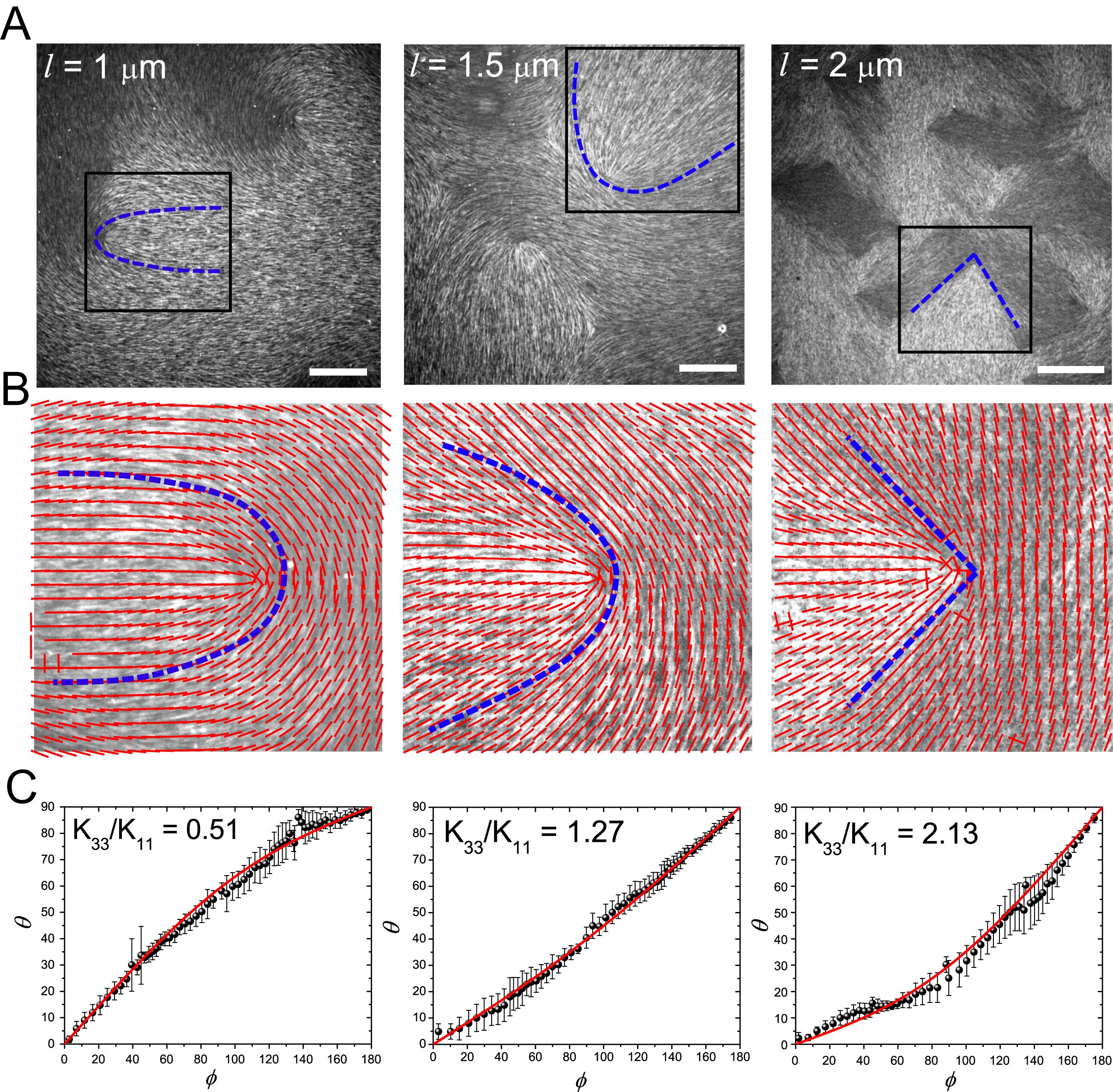}}
%\centering
%\includegraphics[width=0.85\linewidth]{fig2.eps}
\caption{\textbf{Experiments on changing and visualizing actin-based LC's elasticity by varying filament length}. {\bf A}. Fluorescent images of F-actin nematic LC on an oil-water interface. Filament length $l$ increases from left to right. Dashed lines highlight the shape of the +1/2 defects (Scale bar, 30 $\mu$m). {\bf B}. Red lines indicate the local director fields near $+1/2$ defects highlighted in A. {\bf C}. Quantitative measurements of the director fields for the above defects in terms of $\theta(\phi)$ plots.}
\label{fig2}
\end{figure*}
We next demonstrate experimentally that it is possible to vary and visualize the material's elasticity by constructing LCs comprised of the biopolymer F-actin with variable filament length. The filaments are polymerized starting from a 2 $\mu$M solution of monomeric actin (G-actin), $20\%$ of which are fluorescently labeled.
The filament length is varied from $\approx$ 1 $\mu m$ to 2 $\mu m$ in the presence of variable concentrations of capping protein (see Methods).
Actin filaments are crowded onto an oil-water interface stabilized by surfactant molecules by using methylcellulose as a depletion agent. Over a duration of 30 - 60 minutes, a dense film of actin filaments forms a nematic phase (see Movie S1), evidenced by the appearance of several $\pm1/2$ defects, reminiscent of the Schlieren texture of a thermotropic LC after a sudden temperature quench\cite{deGennes_book}. Fig.~\ref{fig2}A shows that as the filamant length is increased from $1~\mu$m to $2~\mu$m, the $+1/2$ defect transitions from a U-shape to a V-shape morphology (see auxiliary dashed lines in Fig.~\ref{fig2}A). The corresponding director field of these images is obtained through image analysis\cite{oakes}, and the structural changes of these defects are shown in Fig.~\ref{fig2}B.
To our knowledge, this is the first experimental realization of such a dramatic transition in defect morphology, and, in fact, is captured quantitatively by the theory described above (Fig.~\ref{fig1}C). According to the Onsager type model \cite{odijk_liqcry1986}, the increase in mesogen length systematically increases $\kappa$ of an LC suppressing bend distortion in the nematic director \cite{Narayan_2006}. This suggests that the underlying mechanism of defect morphogenesis is due to the change in LC's elastic constants.
As we will elaborate further, the agreement between Fig.~\ref{fig1}D and Fig.~\ref{fig2} demonstrates that the nematic elasticity theory can describe polydisperse, semiflexible biopolymer suspensions.
Our experiments also demonstrate that the mechanical properties of actin-based nematic LCs are highly tunable through biochemical control of semiflexible polymer length.

To quantify the LC mechanics, we plot the angle $\theta$ against $\phi$ and extract $\kappa$ by fitting it to the theory.
The calculations shown in Fig.~\ref{fig2}C indicate that actin LCs exhibit $\kappa$ from $0.5$ to $2.1$. Importantly, and in contrast to the behavior of traditional small-molecule thermotropic LCs, one can access both the calamitic regime ($K_{33}>K_{11}$) and the chromonic regime ($K_{33}<K_{11}$) with the same material by using different mesogen lengths.
According to excluded volume theory for rigid-rod LCs, $K_{11}\propto \phi l$ and $K_{33}\propto \phi^3 l^3$, where $l$ is the rod length and $\phi$ is the volume fraction\cite{odijk_liqcry1986,zhou2014elasticity}. Therefore $\kappa \propto \phi^2 l^2$. Our experiments are consistent with this scaling law: as the average filament length is doubled, $\kappa$ increases by a factor of approximately $4.2$.
F-actin is a semiflexible biopolymer, with persistence length $l_p \approx 17~\mu m$ \cite{ott_pre1993}.
Thus, as polymer length increases, one should expect deviations due to their semiflexible nature.  For polymer length $< 2~\mu m$, we find good agreement with rigid-rod theory.  However, for lengths $>2 ~\mu m$, the semiflexible nature becomes increasingly important as $\kappa$ ceases to increase.
%Because the theory is based on perfectly rigid rods, one should expect long filament systems to exhibit deviations due to their finite flexibility. Surprisingly, the deviation of $\kappa$ in our experiments is negligible, suggesting that the actin filaments as long as 2~$\mu$m act as ideal rigid rods.
%Interestingly, we find that although the dependence of $\kappa$ on $l$ is consistent with the Onsager type model, its prediction of $\kappa$'s absolute value is not, possibly due to the semiflexibility of the actin filaments, similar to that observed in lyotropic chromonic LCs \cite{zhou2014elasticity}.

\begin{figure*}%[tbhp]
 \centerline{\includegraphics[width=0.85\textwidth]{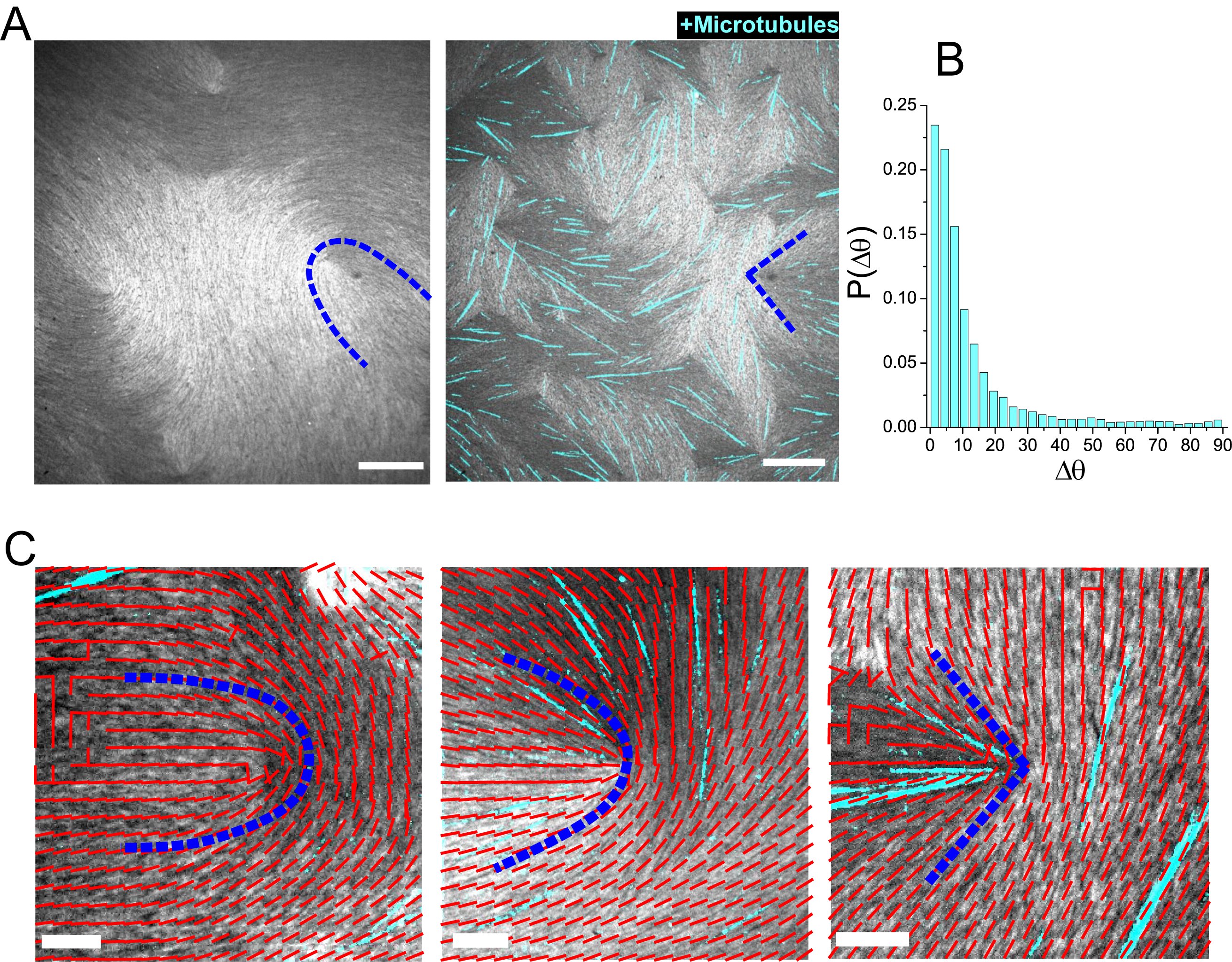}}
%\centering
%\includegraphics[width=0.85\linewidth]{fig3.eps}
\caption{ \textbf{Experiments on changing actin-based LC's elasticity by adding rigid microtubules}. {\bf A}. Optical images of actin LC without and with microtubules. Blue dashed lines highlight the change in defect shape from U to V. (Scale bar, 30 $\mu$m) {\bf B}. Probability distribution of the angle between microtubule orientations and the local F-actin director fields. {\bf C}. Optical images of $+1/2$ defects overlaid with the corresponding director field. Microtubule number density increases from left to right followed by the defect shape change from U to V. (Scale bar, 10 $\mu$m).}
\label{fig3}
\end{figure*}
When the filament length is comparable to its persistence length, the bend-to-splay ratio $\kappa$ ceases to increase and instead declines (see Fig. \ref{longactin} in the SI), consistent with previous work in poly-$\gamma$-benzylglutamate based lyotropic LC\cite{lee_liqcry1990,lee_meyer_book}. Therefore, there is limitation on varying the LC's elasticity by solely relying on changing the length of a semiflexible mesogen.

To explore the possibility of accessing a larger range of LC elasticity, we created composite LCs consisting of concentrated actin suspensions inter-dispersed with sparse, more rigid, biopolymers, microtubules ($l_p \approx$ 1 mm). Intuitively, microtubules, if sufficiently sparse, should not affect the splay constant, but should increase the bend elasticity of the LC.
We explore this possibility by constructing a simple model, based on three assumptions: (1) microtubules are well dispersed in the actin, (2) microtubules are relatively short and only weakly bent in the LC and (3) microtubule density is sufficiently small to render self-interactions negligible.
We observe that microtubules are aligned with the local director field (Fig.~\ref{fig3}B). This substantiates our intuition that the rigid polymers should only suppress the bend mode (increase $K_{33}$) without penalizing splay mode ($K_{11}$ unchanged). The free energy penalty associated with microtubule bending can be incorporated into the bend distortion term $f_{bend}$ in the Frank-Oseen expression, such that the change in bend constant $\Delta K_{33}$ is a linear function of microtubule number density $c$ (see Supplementary Information for the details of derivation):
\begin{equation}\label{deltaK2d}
\Delta K_{33}=\frac{E I}{\delta_z}  cl_0,
\end{equation}
where $l_0$ is the average microtubule length, $EI\approx 4.1\times10^{-24}~ J\cdot m$ is the microtubule flexural rigidity\cite{blanchoin2014actin,hawkins_biophys2013}, and $\delta_z$ is the film thickness. Our measurements estimate $\delta_z\approx300~nm$ (see Fig. \ref{thickness} in the SI).

Guided by the considerations outlined above, we add a small concentration of taxol-stabilized microtubules to the solution of actin filaments.
Consistent with our hypothesis, we observe that the addition of microtubules has a significant effect on defect morphology. Figure~\ref{fig3}A shows optical images of a system with and without microtubules. In the absence of microtubules, the +1/2 defects' morphology is U-shaped, as outlined by blue dashed line and a significant amount of bending is also visible. When microtubules are added, all defects adopt a V-shape morphology, and bending is suppressed. In Fig.~\ref{fig3}C, we find that the number of microtubules, shown in cyan, has a direct influence on the shape of +1/2 defects, which undergoes a gradual transition from a U- to a V-shape morphology, as highlighted by the director field around the defect core. Eq.~\ref{deltaK2d} indicates that the change in the composite LC's elasticity should only be a function of $cl_0$, which is namely microtubule's linear density.
 \begin{figure*}
  \centerline{\includegraphics[width=0.85\textwidth]{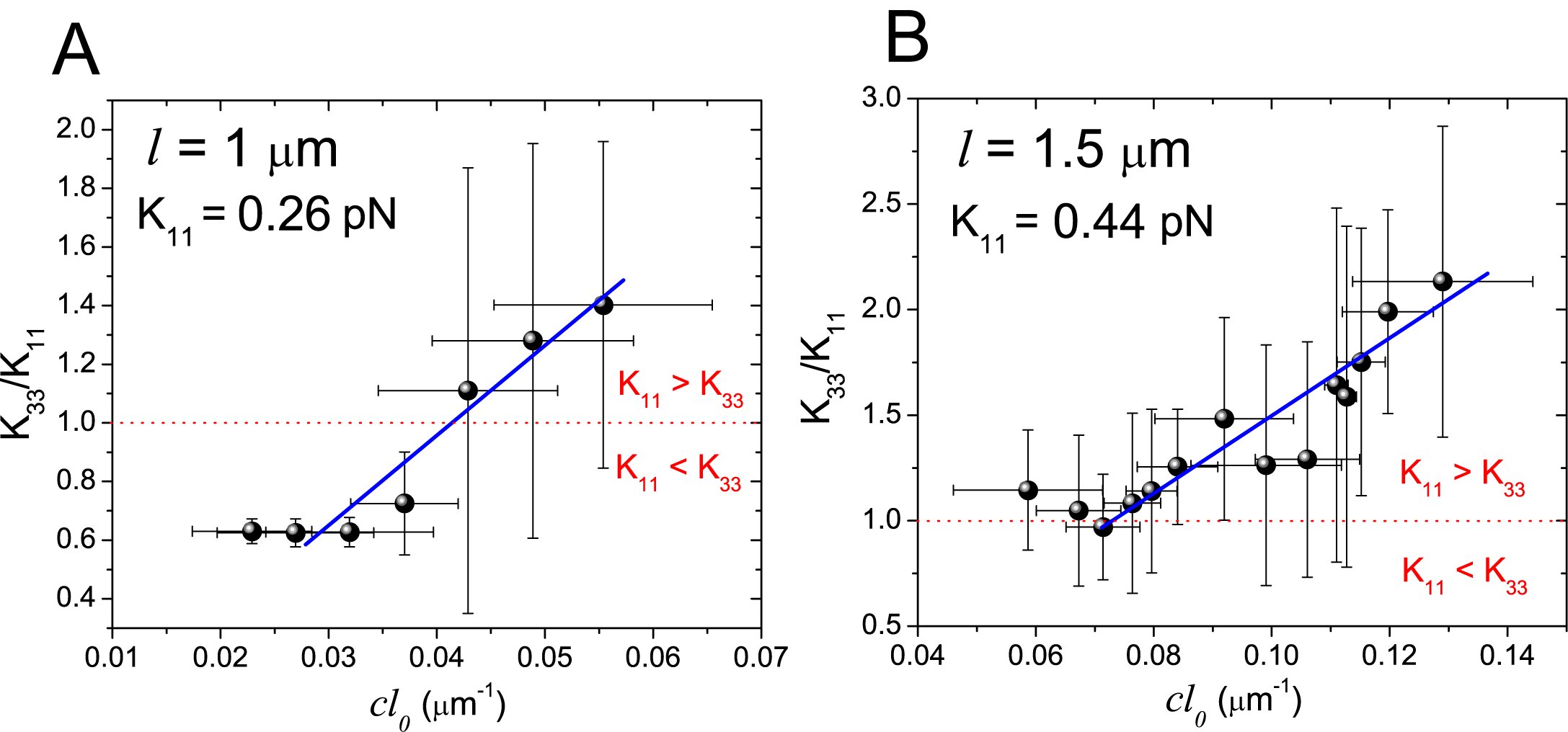}}
%\centering
%\includegraphics[width=0.85\linewidth]{fig4.eps}
\caption{ \textbf{Experiments on changing actin-based LC's elasticity by adding rigid microtubules}. The two plots are two independent experimental sets, each of which plots the fitted elastic constant ratio $\kappa $ as a function of $cl_0$ for {\bf A}. $l$ = 1 $\mu$m for $15$ defects and {\bf B}. $l$ = 1.5 $\mu$m for 20 defects. The slope gives the direct measure of the splay elastic constant $K_{11}$. The error bars correspond to the standard deviation.}
\label{fig4}
\end{figure*}

To test this prediction, we inspect several tens of $+1/2$ defects for two independent experimental sets. The microtubule's linear density is determined by measuring $c$ and $l_{0}$ in the vicinity of each individual defect (see Methods). We measure $\kappa$ by fitting the defect morphology to the theory, and we then plot it against $cl_0$. In Fig.~\ref{fig4} we find that both independent sets of experiments show a linear trend between $\kappa $ and $cl_0$ above a threshold of microtubule's linear density. In Fig.~\ref{fig4}A, the transition from a splay-dominated into a bend-dominated regime that occurs at $K_{33}=K_{11}$ is observed, exhibiting the high tunability of the composite LC.

Having validated the theory, this framework can be used to measure the absolute values of $K_{11}$ and $K_{33}$. Since the elastic constant ratio $\kappa$ is measured before and after microtubules are added, $\Delta \kappa = \Delta K_{33}/K_{11}$ is known.
Using Eq.~\ref{deltaK2d} to determine $\Delta K_{33}$, $K_{11}$ can be evaluated from $ \Delta K_{33}/\Delta \kappa$. Our calculations indicate that for actin LCs  with $l=1.5~\mu m$, $K_{11}=K_{33}\approx 0.44~$pN, and for actin LCs with $l=1~\mu m$, $K_{11}\approx 0.26~$pN and $K_{33}\approx0.16~$pN. The fact that the long-filament system has a higher elastic constant and higher $\kappa$ is also consistent with the Onsager type model. Note that in the limit of high concentration of microtubules, the composite's properties will deviate from the linear equation and approach those of a pure microtubule LC.
%The threshold linear density $cl_0$ beyond which the theory ceases to work (see Fig.~\ref{fig4}) may be attributed to the fact that many microtubules are not well engaged with the actin filaments. They instead stack on top of the film of actin LC, and have little effect on the LC's elasticity, but are still visible in confocal microscopy.

Traditional LC elasticity measurements rely on external fields or calibrations \cite{deGennes_book,zhou_prl2012,zhou2014elasticity}, a limitation overcome by the use of this protocol.
Microtubules in the system play the role of ``elastic dopants'', analogous to the chiral dopants that are added to nematic materials to increase chirality\cite{xu_pre1997}. Our results provide a framework within which one can simultaneously visualize, tune and measure the material's elasticity. We emphasize here that the idea of measuring material's mechanical properties by observing defects can be generalized to other ordered systems. Defects are singularity regions in an otherwise ordered material, where different elastic modes compete against each other. 
Thus, we expect the framework developed here could be extended to characterization of the morphology and statistics of defects, disclinations, or dislocations as a means to probe the elastic properties of other types of materials. 

\begin{figure*}
 \centerline{\includegraphics[width=0.9\textwidth]{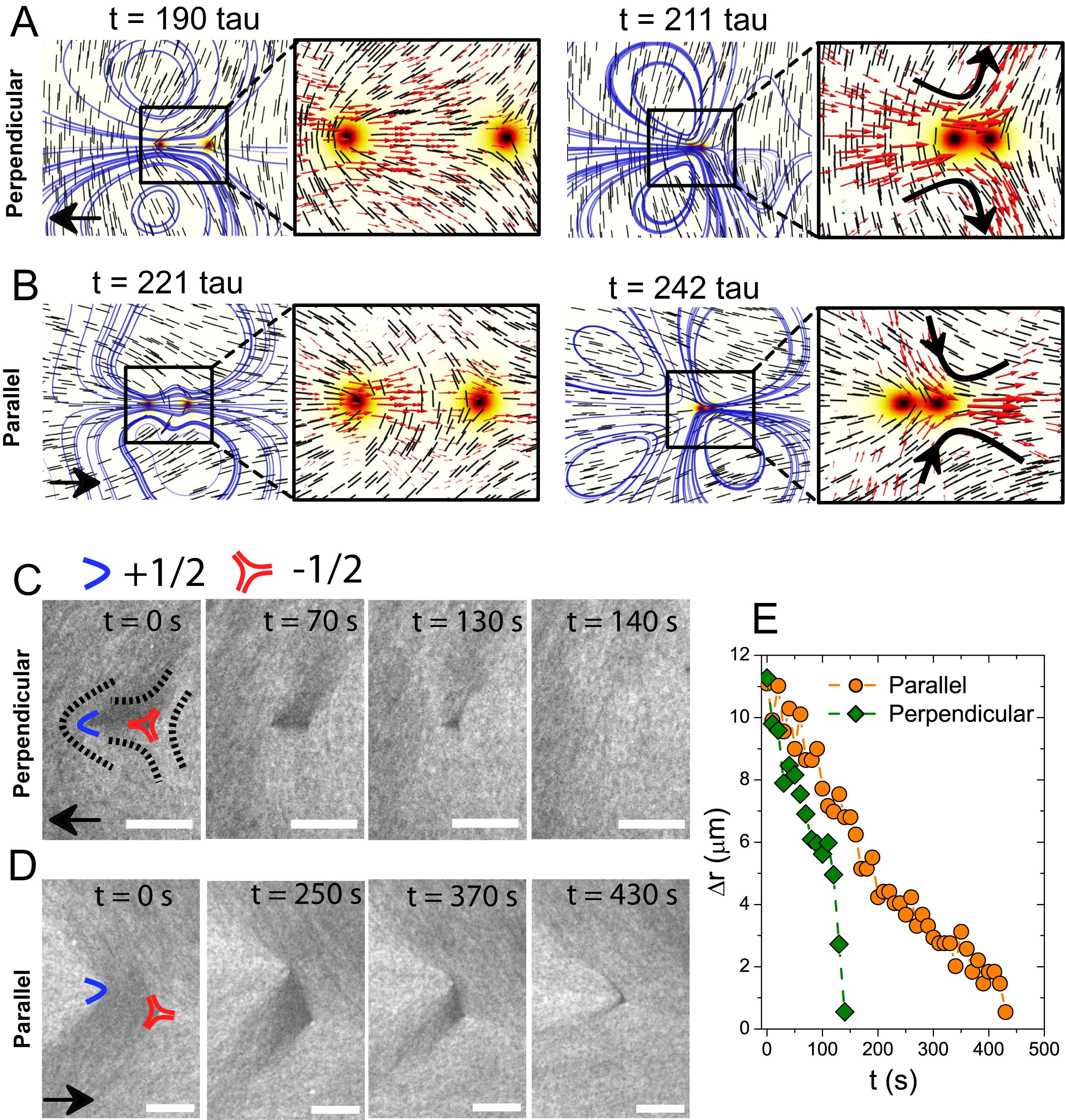}}
%\centering
%\includegraphics[width=0.9\linewidth]{fig5.eps}
\caption{\textbf{Two scenarios of defect annihilation events}. {\bf A} \& {\bf B}. Simulations of defect annihilation with two different defect configurations; black arrows indicate the orientations of $+1/2$ defects approaching -1/2 defect from left; Short lines show the director field and curves are the streamlines. Background color corresponds to the order parameter using the same colorbar as in Fig. \ref{fig1}D. In zoom-in frames, black arrows indicate the inward/outward flows upon annihilation, and red arrows represent the velocity field. {\bf C} \& {\bf D}. Time sequence of annihilation as seen in the experiments for perpendicular and parallel case, respectively. Scale bar equals 10 $\mu$m in both \textbf{C} and \textbf{D}. Blue boomerangs and red triangles represent +1/2 and -1/2 defects, respectively. \textbf{E}. Defect separation, $\Delta r$, plotted as function of time.}
\label{fig5}
\end{figure*}

The results presented thus far serve to establish that the theoretical framework adopted here can in fact describe the equilibrium structure of lyotropic biopolymers. For the applications of active lyotropic materials that we envisage, however, it is essential that such a formalism be also able to describe the transport of molecular cargo in topological defects. In what follows, we therefore turn our attention to the more challenging task of characterizing and predicting the coupling between structure and dynamics that arises in actin based liquid crystals.  When a thermotropic LC is quenched from an isotropic into a nematic state, defects emerge and annihilate each other over time. The interplay between LC elasticity, viscosity, and defect structure (configurations), determines the dynamics of annihilation events. Note that recent efforts have addressed the coupling of structure and dynamics in free-standing smectic LCs\cite{stannarius_prl2016}. In contrast to that work, our actin-based LC system represents a true nematic LC which is free of confining walls. As such, it offers a unique 2D platform on which to explore structure-based dynamics.

In Fig.~\ref{fig5}, we show two typical defect configurations, which differ by the angle between the line connecting the defect cores and the nematic far field. For simplicity, and without loss of generality, we consider two opposite, symmetric cases. In the first, the defect line is parallel to a well-defined far field; in the second, it is perpendicular (Fig.~\ref{fig5}A-B).
In both situations, the defects attract each other in order to lower the system's total elastic free energy. Thus in the ``perpendicular'' case, the $+1/2$ defect should move opposite to its orientation; whereas in the ``parallel'' case, the $+1/2$ defect should move along its orientation (See supplementary movie S2).
This is exactly what is observed in our experiments, as shown in Fig.~\ref{fig5}C and \ref{fig5}D for the perpendicular and parallel cases, respectively and supplementary movie S3.
Interestingly, in the case of the parallel far field (Fig.~\ref{fig5}D), the +1/2 defect deviates from a straight head-on approach to the -1/2 defect. 
To the best of our knowledge, this result provides a first direct visualization of two distinct defect annihilations in nematic LCs.

We use our theoretical model to investigate how the flow field generated by the defect motion depends on the LC's elastic constants and the underlying configuration.
When two $\pm 1/2$ defects approach, and eventually annihilate each other, hydrodynamic effects play a key role in the dynamics. In particular, past work has suggested that the $+1/2$ defect always moves faster than the $-1/2$ defect\cite{toth2002hydrodynamics}. Our hydrodynamic simulations and experimental observations, however, indicate that this is not the complete story. When one takes into account elasticity and defect configurations, the opposite can be true: a $-1/2$ defect may move faster than a $+1/2$ defect (see Supplementary Information for a detailed discussion). In Fig.~\ref{fig5}, we show the effect of defect configurations on an isolated annihilation event. We find that the ``perpendicular'' case annihilates faster than the ``parallel'' case. If one ignores hydrodynamic effects, the two cases would proceed in an identical manner (under the one-constant-assumption). The inclusion of hydrodynamics breaks that symmetry. As shown in Fig.~\ref{fig5}, the flow fields corresponding to the two configurations are markedly different.  For the perpendicular case, there are two symmetric vortices on the two sides of the defect line (indicated by blue lines in Fig.~\ref{fig5}A with an outward transverse flow upon annihilation). Whereas for the parallel case, upon defect annihilation, an extensional flow is produced, and the transverse flow is inward. Because the director field at the defect line in the ``perpendicular'' case is aligned with the flow, the corresponding viscosity is low and the enhanced hydrodynamic flow accelerates the defects' motion. In the ``parallel'' case, the flow direction is perpendicular to the local director field, and the high viscosity suppresses such flow, leading to slower dynamics. The horizontal separation of two $\pm1/2$ defects, $\Delta$r, can be plotted as a function of time as shown in Fig.~\ref{fig5}E. Our experiments indicate that on average, the ``perpendicular'' case clearly exhibits faster dynamics than the ``parallel'' case, confirming our predictions. Taken together, these results establish that defect structure influences dynamics, a feature that could be exploited to manipulate the transport of cargo in an LC by creating different flow patterns as defects annihilate.

\begin{figure*}
 \centerline{\includegraphics[width=1\textwidth]{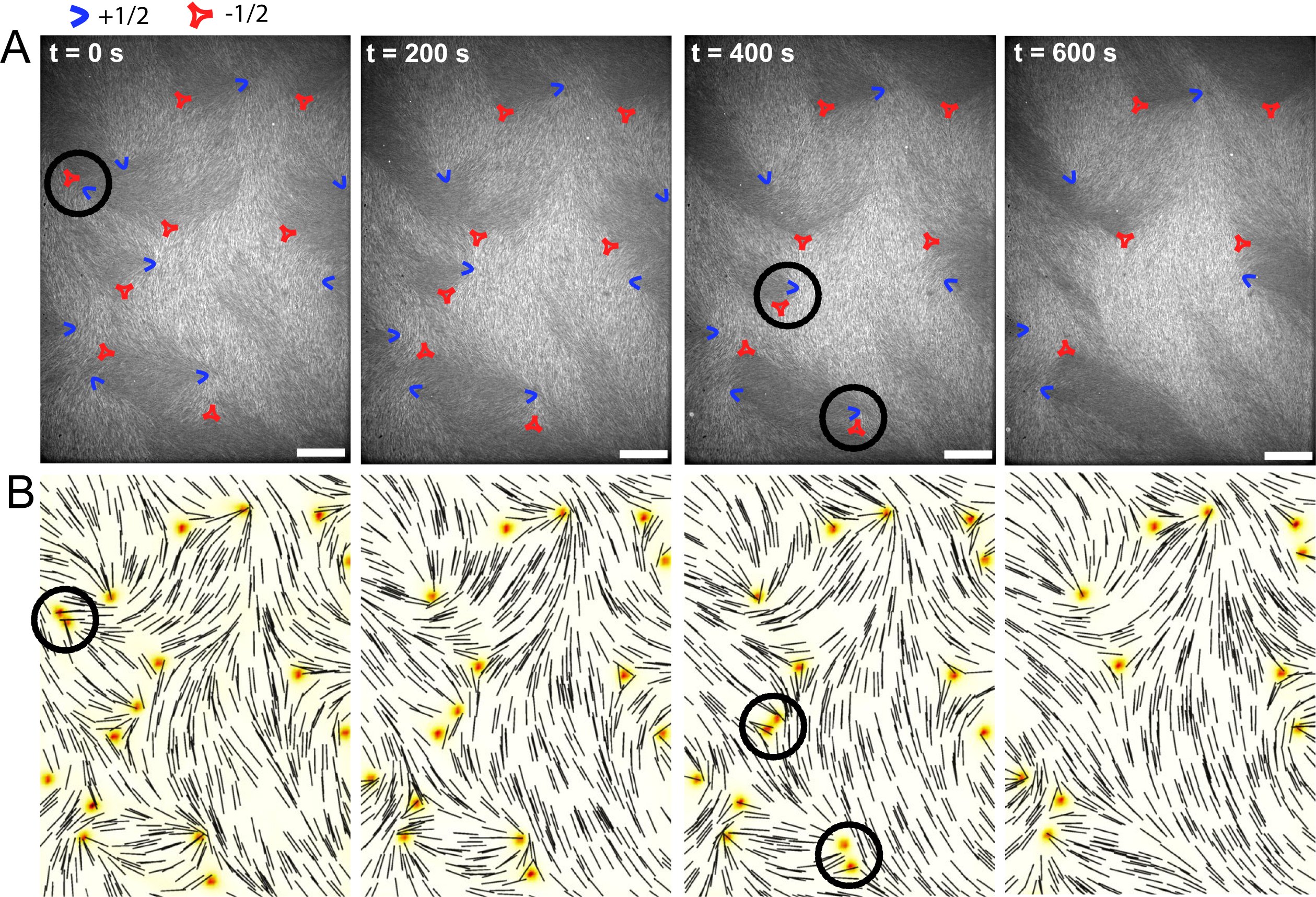}}
%\centering
%\includegraphics[width=1\linewidth]{fig6.eps}
\caption{ \textbf{Actin-based LC ``weather map''}. Top row: consecutive optical images from actin LC equilibration process; blue boomerangs represent $+1/2$ defects and red triangles indicate $-1/2$ defects. Bottom row: consecutive simulation images corresponding to the experiments; Black lines depict the local director field and the background color indicates scalar order parameter (adopting the same colorbar as in Fig.~\ref{fig1}). The circled defects annihilate in the next frame. Scale bar is $30~\mu m$.}
\label{fig6}
\end{figure*}

The results presented in Fig.~\ref{fig5} are limited to two isolated defects and only two ideal, extreme cases (parallel and perpendicular defect configuration) are considered. The question that emerges then is whether the simple formalism adopted here can describe dynamics in the more common case in which multiple defects with nearly arbitrary relative orientations interact with each other\cite{vromans_softmatt2016}.
To address this question, we now consider a system of area $\approx 10^5 ~\mu m^2$ containing tens of defects. We measure the director field from the experiments and use it as the initial condition for a 3D simulation (see Methods). We run the simulation over an extended period of time, and examine whether the model can capture the temporal evolution of defects that is observed in the experiments over laboratory time scales (minutes to hours).
Our computer simulation can be viewed as analogous to weather forecasting: we take a pattern (director field) as the initial condition, and we rely on a simulation to predict its structure at a future time. In our quasi-2D nematic system, we show four images comparing simulations and experiments at four consecutive times, separated by a time interval of approximately 200 seconds. As shown in Fig. 6 and supplementary movie S4, the model correctly predicts the three annihilation events (see the circled defects in Fig.~\ref{fig6}), and individual defect positions after 10 minutes.
This agreement, which is simply based on the initial defect structure, serves to underscore that the hydrodynamic formalism adopted here is able to describe the structure and dynamics of lyotropic, polydisperse, actin-based 2D nematic systems. Despite the fact that the LC's elastic structure and hydrodynamics are both long-ranged and nonlinearly coupled, its defect dynamics can be accurately predicted by our athermal model, implying that such defect dynamics are deterministic.

In conclusion, we have demonstrated a semiflexible biopolymer-based 2D nematic system, which exhibits highly tunable elasticity, and could be useful for future LC technologies in compartmentalization and transport of molecular cargo. Its elastic constants can be tuned by either varying the average filament length, or by doping it with a sparse concentration of rigid polymers. We show a method to infer the ratio of splay and bend constant by visualizing the morphology of $+1/2$ defects. Using this, we find that the bend-to-splay ratio can range from $0.5$ to $2.1$, such that the material transitions from a chromonic-like to a calamatic-like material. For rigid-polymer doped composites, our simple theoretical model predicts that the change in bend constant is proportional to the linear density of rigid polymers, and enables measurement of the absolute values of bend and splay constants.
We further show that the hydrodynamic model of nematic liquid crystals developed in our work describes defect dynamics of this highly structured viscoelastic fluid over extended periods of time. In particular, we have predicted and confirmed experimentally that the annihilation time of a $\pm 1/2$ defect pair depends on the defect configuration, a feature that could potentially be used to engineer molecular transport systems that rely on autonomous defect motion.

\subsection*{Theory and modeling}
The bulk free energy of the nematic LC, $F$, is defined as
\begin{equation}\label{total}
\begin{aligned}
F&= \int_{V} dV f_{bulk} + \int_{\partial V} dS f_{surf} \\
&= \int_{V} dV (f_{LdG}+f_{el}) + \int_{\partial V} dS f_{surf},
\end{aligned}
\end{equation}
where $f_{LdG}$ is the short-range free energy, $f_{el}$ is the long-range elastic energy, and $f_{surf}$ is the surface free energy due to anchoring. $f_{LdG}$ is given by a Landau-de Gennes expression of the form\cite{deGennes_book,landau_statistics_book}:
\begin{equation}\label{phase}
f_{LdG}=\frac{A_0}{2}(1-\frac{U}{3})\tr({\bf Q}^2)-\frac{A_0U}{3}\tr({\bf Q}^3)+\frac{A_0 U}{4}(\tr({\bf Q}^2))^2.
\end{equation}
Parameter $U$ controls the magnitude of $q_0$, namely the equilibrium scalar order parameter via $q_0=\frac{1}{4}+\frac{3}{4}\sqrt{1-\frac{8}{3U}}$.
The elastic energy $f_{el}$ is written as ($Q_{ij,k}$ means $\partial_k Q_{ij}$):
\begin{equation}
\label{elastic_en}
\begin{aligned}
f_{el}=&\frac{1}{2}L_1 Q_{ij,k}Q_{ij,k}+\frac{1}{2}L_2 Q_{jk,k}Q_{jl,l}\\
&+\frac{1}{2}L_3 Q_{ij}Q_{kl,i}Q_{kl,j}+\frac{1}{2}L_4 Q_{ik,l}Q_{jl,k}.
\end{aligned}
\end{equation}
If the system is uniaxial, the $L$'s in Eq.~\ref{elastic_en} can be mapped to the $K$'s in Eq.~\ref{frank} via
\begin{equation}
\begin{aligned}
L_1&=\frac{1}{2q_0^2} \left[ K_{22}+\frac{1}{3}(K_{33}-K_{11})  \right], \\
L_2&=\frac{1}{q_0^2} (K_{11}-K_{24}), \\
L_3&=\frac{1}{2q_0^3} (K_{33}-K_{11}), \\
L_4&=\frac{1}{q_0^2} (K_{24}-K_{22}).
\end{aligned}
\end{equation}
By assuming a one-elastic-constant $K_{11}=K_{22}=K_{33}=K_{24}\equiv K$, one has $L_{1}=L\equiv K/2q_0^2$ and $L_{2}=L_{3}=L_{4}=0$.
Point wise, ${\bf n}$ is the eigenvector associated with the greatest eigenvalue of the ${\bf Q}$-tensor at each lattice point.

To obtain the static morphology of topological defects, we minimize Eq.~\ref{total} with respect to the {\bf Q}-tensor via a Ginzburg-Landau algorithm\cite{zumer_ldG_review}. To simulate the LC's non-equilibrium dynamics, a hybrid lattice Boltzmann method is used to simultaneously solve a Beris-Edwards equation and a momentum equation which accounts for the back-flow effects.
By introducing a velocity gradient $W_{ij}=\partial_j u_i$, strain rate $\bf A=(\bf W + \bf W^T)/2$, vorticity $\bf \Omega=(\bf W - \bf W^T)/2$, and a generalized advection term
\begin{equation}
\begin{aligned}
{\bf S}({\bf W},{\bf Q})=&(\xi {\bf A}+{\bf \Omega})({\bf Q}+{\bf I}/3)+({\bf Q}+{\bf I}/3)(\xi {\bf A}-{\bf \Omega})&\\
&-2\xi ({\bf Q}+{\bf I}/3) \tr({\bf QW}),&
\end{aligned}
\end{equation}
one can write the Beris-Edwards equation\cite{beris_edwards_book} according to
\begin{equation} \label{beris_edwards_eq}
(\partial_t +{\bf u}\cdot \nabla){\bf Q}-{\bf S}({\bf W},{\bf Q})=\Gamma \bf{H}.
\end{equation}
The constant $\xi$ is related to the material's aspect ratio, and $\Gamma$ is related to the rotational viscosity $\gamma_1$ of the system by $\Gamma=2q_0^2/\gamma_1$\cite{denniston_2d}. The molecular field $\bf{H}$, which drives the system towards thermodynamic equilibrium, is given by
\begin{equation}
{\bf H}=-\left[ \frac{\delta F}{ \delta \bf{Q}} \right]^{st},
\end{equation}
where $\left[ ...\right]^{st}$ is a symmetric and traceless operator. When velocity is absent, i.e. ${\bf u}({\bf r})\equiv 0$, Besris-Edwards equation Eq.~\ref{beris_edwards_eq} reduce to Ginzburg-Landau equation:
$$
\partial_t {\bf Q}=\Gamma {\bf H}.
$$
To calculate the static structures of $\pm 1/2$ defects, we adopt the above equation to solve for the ${\bf Q}$-tensor at equilibrium.

Degenerate planar anchoring is implemented through a Fournier-Galatola expression\cite{fournier_galatola} that penalizes out-of-plane distortions of the ${\bf Q}$ tensor. The associated free energy expression is given by
\begin{equation}
f_{surf} = W (\tilde{\bf Q} - \tilde{\bf Q}^{\perp})^2,
\end{equation}
where $\tilde{{\bf Q}} = {\bf Q} + (q_0/3)\bf{I}$ and $\tilde{\bf{Q}}^{\perp} = {\bf P} \tilde{\bf{Q}} {\bf P}$. Here ${\bf P}$ is the projection operator associated with the surface normal ${\bf \nu}$ as ${\bf P}=\bf{I} -{\bf \nu} {\bf \nu} $. The evolution of the surface ${\bf Q}$-field is governed by\cite{zhang_jcp2016}:
\begin{equation} \label{surface_evolution}
\frac{\partial {\bf Q}}{\partial t}=-\Gamma_s \left( -L {\bf \nu} \cdot \nabla {\bf Q} +\left[  \frac{\partial f_{surf}}{\partial {\bf Q}} \right]^{st} \right),
\end{equation}
where $\Gamma_s=\Gamma/\xi_N$ with $\xi_N=\sqrt{L_1/A_0}$, namely nematic coherence length. The above equation is equivalent to the mixed boundary condition given in Ref.~\cite{dagama_lblc} for steady flows.

Using an Einstein summation rule, the momentum equation for the nematics can be written as\cite{denniston_2d,denniston_3d}
\begin{equation}  \label{ns_eq}
\begin{aligned}
\rho(\partial_t+u_\beta \partial_\beta)u_\alpha=&\partial_\beta \Pi_{\alpha_\beta}+\eta\partial_\beta[\partial_\alpha u_\beta+\partial_\beta u_\alpha\\
&+(1-3\partial_\rho P_0)\partial_\gamma u_\gamma \delta_{\alpha_\beta}].
\end{aligned}
\end{equation}
The stress ${\bf \Pi}$ is defined as
\begin{equation}
\begin{aligned}\label{stress}
\Pi_{\alpha \beta}=  & -P_0 \delta_{\alpha \beta}-\xi H_{\alpha \gamma} (Q_{\gamma \beta} +\frac{1}{3}\delta_{\gamma \beta}) - \xi (Q_{\alpha \gamma} +\frac{1}{3} \delta_{\gamma \beta} ) H_{\gamma \beta} & \\
 &+ 2 \xi (Q_{\alpha \beta} +\frac{1}{3}\delta_{\alpha \beta}) Q_{\gamma \epsilon}H_{\gamma \epsilon} -\partial_{\beta} Q_{\gamma \epsilon} \frac{\delta \mathcal F}{\delta \partial_\alpha Q_{\gamma \epsilon}} &\\
 & + Q_{\alpha \gamma} H_{\gamma \beta} -H_{\alpha \gamma} Q_{\gamma \beta}, &
\end{aligned}
\end{equation}
where $\eta$ is the isotropic viscosity, and the hydrostatic pressure $P_0$ is given by\cite{fukuda_force}
\begin{equation}
P_0=\rho T - f _{bulk}.
\end{equation}
The temperature $T$ is related to the speed of sound $c_s$ by $T=c_s^2$.
We solve the evolution equation Eq.~\ref{beris_edwards_eq} using a finite-difference method. The momentum equation Eq.~\ref{ns_eq} is solved simultaneously via a lattice Boltzmann method over a D3Q15 grid\cite{zhaoliguo_book}. The implementation of stress follows the approach proposed by Guo {\it et al.}\cite{zhaoliguo_forcing}. Our model and implementation were validated by comparing our simulation results to predictions using the ELP theory\cite{ELP_ericksen, ELP_leslie, ELP_parodi,lavrentovich_book}.
The units are chosen as follows: the unit length $a$ is chosen to be $a=\xi_N=1~\mu m$, characteristic of the filament length, the characteristic viscosity is set to $\gamma_1=0.1~Pa\cdot s$, and the force scale is made to be $F_0=10^{-11}$~N. $L_1=0.0222$, $L_3=0.0428$, and $L_2=L_4=0$ lead to $\kappa\equiv K_{33}/K_{11}=3$ and $K_{11}=K_{22}=K_{24}$. Other parameters are chosen to be $A_0=0.132$, $\xi=0.8$, $\Gamma=0.2$, $\eta=0.27$, and $U=3.5$ leading to $q_0\approx 0.62$.
The simulation is performed in a rectangular box. The boundary conditions in the $xy$ plane are periodic. Two confining walls are introduced in the $z$ dimension, with strong degenerate planar anchoring, ensuring a quasi 2D system. In the ``weather map'' simulation, $L_x=300~\mu m$, $L_y=200~\mu m$, slightly larger than the experimental image, in order to suppress boundary effects; $L_z=7 ~\mu m$ is chosen to correspond to the sample thickness, mimicking the hydrodynamic screening effects of the substrates in the experiments.
We refer the reader to Ref.~\cite{zhang_jcp2016} for additional details on the numerical methods employed here.

\subsection*{Experimental methods}
\subsubsection*{Proteins}
Monomeric actin is purified from rabbit skeletal muscle acetone powder \cite{actin} (Pel-Freez Biologicals, Rogers, AR) stored at $-80^0$C in G-buffer(2mM Tris HCL pH 8.0, 0.2 mM APT, 0.2 mM CaCl$_{2}$, 0.2 mM DTT, $0.005\%$ NaN$_{3}$). Fluorescent labeling of actin was done with a tetramethylrhodamine-6-maleimide dte (Life Technologies, Carlsbad, CA). Capping protein [mouse, with a HisTag, purified from bacteria \cite{CP}; gift from the Dave Kovar lab, The University of Chicago, Chicago, IL] is used to regulate actin filament length. Mictotubules are polymerized in PEM-100 buffer at 37$^\circ$ C (100mM Na-PIPES, 1mM MgSO$_{4}$, 1mM EGTA, pH 6.8) in the 1:10 ratio of fluorescently labeled tubulin (Cytoskeleton, cat$\#$ TL488M) and unlabeled tubulin (Cytoskeleton, cat$\#$ HTS03) in the presence of 1 mM GMPCPP (Jena Biosciences, cat $\#$ NU-405L). They are later stabalized by adding 50 $\mu$M Taxol. The microtubule length is shortened by shearing through Hamilton Syringe (Mfr $\#$ 81030, Item $\#$ EW-07939-13) before adding to the actin polymers.

\subsubsection*{Experimental assay and microscopy}
The assay consists of 1X F-buffer (10 mM imidazole, pH 7.5, 50mM KCL, 0.2mM EGTA, 1mM MgCl$_{2}$ and 1mM ATP). To prevent photobleaching, an oxygen scavenging system (4.5 mg/mL glucose, 2.7 mg/mL glucose oxidase(cat$\#$345486, Calbiochem, Billerica, MA), 17000 units/mL catalase (cat $\#$02071, Sigma, St. Louis, MO) and 0.5 vol. $\%$ $\beta$-mercaptaethanol is added. 0.3 wt $\%$ 15 cP methylcellulose is added as the crowding agent.

Actin from frozen stocks stored in G-buffer is added to a final concentration of 2 $\mu$M with a ratio 1:5 TMR-maleimide labeled:unlabeled actin monomer. To control the filament length, frozen capping protein stocks are thawed on ice and are added at the same time (0.3 mol $\%$  $\textendash$ 0.17 mol $\%$).

The experiment is done in a cylindrical chamber which is a glass cylinder glued to a coverslip. Coverslips are sonicated clean with water and ethanol. The surface is treated with triethoxy(octyl)silane in isopropanol to produce a hydrophobic surface. To prepare a stable oil-water interface, PFPE-PEG-PFPE surfactant (cat $\#$ 008, RAN biotechnologies, Beverly, MA) is dissolved in Novec-7500 Engineered Fluid (3M, St Paul, MN) to a concentration of 2 wt. $\%$. To avoid flows at the surface, a small $2\times2$ mm teflon mask is placed on the treated coverslip before exposing it to UV/ozone for 10 minutes. The glass cylinder is thoroughly cleaned with water and ethanol before gluing it to the coverslip using instant epoxy. Then $3~\mu$L of oil-surfactant solution is added into the chamber, and quickly pipetted out to leave a thin coating. Actin polymerization mixture was immediately added afterwards. Composite liquid crystals were formed by adding taxol-stabilized microtubules to 1 $\mu g/ml$ final concentration.

The sample is imaged using an inverted microscope (Eclipse Ti-E; Nikon, Melville, NY) with a spinning disk confocal head (CSU-X, Yokagawa Electric, Musashino, Tokyo, Japan), equipped with a CMOS camera (Zyla-4.2 USB 3; Andor, Belfast, UK). A 40X 1.15 NA water-immersion objective (Apo LWD; Nikon) was used for imaging. Images were collected using 568 nm and 647 nm excitation for actin and microtubules respectively. Image acquisition was controlled by Metamorph (Molecular Devices, Sunnyvale, CA).

\subsubsection*{Image Analysis}
The microtubule number density ($c$) and average length ($l_{0}$) is estimated using the ImageJ software \cite{imagej}. The image around a +1/2 defect is cropped to a rectangle shape of area $A$ centered at its core (Fig. \ref{fig3}C).
%Since the microtubules are homogeneously dispersed, both $c$ and $l_{0}$ are independent of the size of the rectangular selection.
The resulting image is then thresholded to separate all the microtubule bundles whose $l_{0}$ is extracted from the major axis of the fitted ellipse. In order to extract $c$, we plot the distribution of background-corrected microtubule intensity and assign the single microtubule intensity ($I_{0}$) to the value where the distribution shows a peak. The bundle intensity $I$ is scaled with $I_{0}$ to extract $n$(=$I/I_{0}$) which is the number of microtubules in a bundle. $c$ is estimated by calculating $\Sigma n/A$, where $\Sigma$ denotes the sum over all the bundles.

\begin{acknowledgments}
This work was supported primarily by the University of Chicago Materials Research Science and Engineering Center, which is funded by the National Science Foundation under award number DMR-1420709. MLG and JR acknowledge support from NSF Grant MCB-1344203. The design of tublin-actin composits in the JR and JJdP group was supported by the U.S. Army Research Office through the MURI program (W911NF-15-1-0568). NK acknowledges the Yen Fellowship of the Institute for Biophysical Dynamics, The University of Chicago. We thank helpful discussions with Dr. Shuang Zhou, Dr. Kimberly Weirich, Dr. Samantha Stam and Dr. Takuya Yanagimachi. 
\end{acknowledgments}

\section*{Author Contribution}
RZ, NK, MLG, and JJdP designed the research, RZ and JJdP developed the model and carried out simulation, NK performed experiments. JR contributed to microtubule experiments. RZ and NK analyzed the data. RZ, NK, MLG, and JJdP wrote the paper.

% \pnasbreak splits and balances the columns before the references.
% If you see unexpected formatting errors, try commenting out this line
% as it can run into problems with floats and footnotes on the final page.
%\pnasbreak

% Bibliography
\bibliography{passive_actin}
\newpage

\section*{Supplementary Information}
\subsubsection*{$-1/2$ defect core shape}

\begin{figure}%[tbhp]
  \centerline{\includegraphics[width=0.85\textwidth]{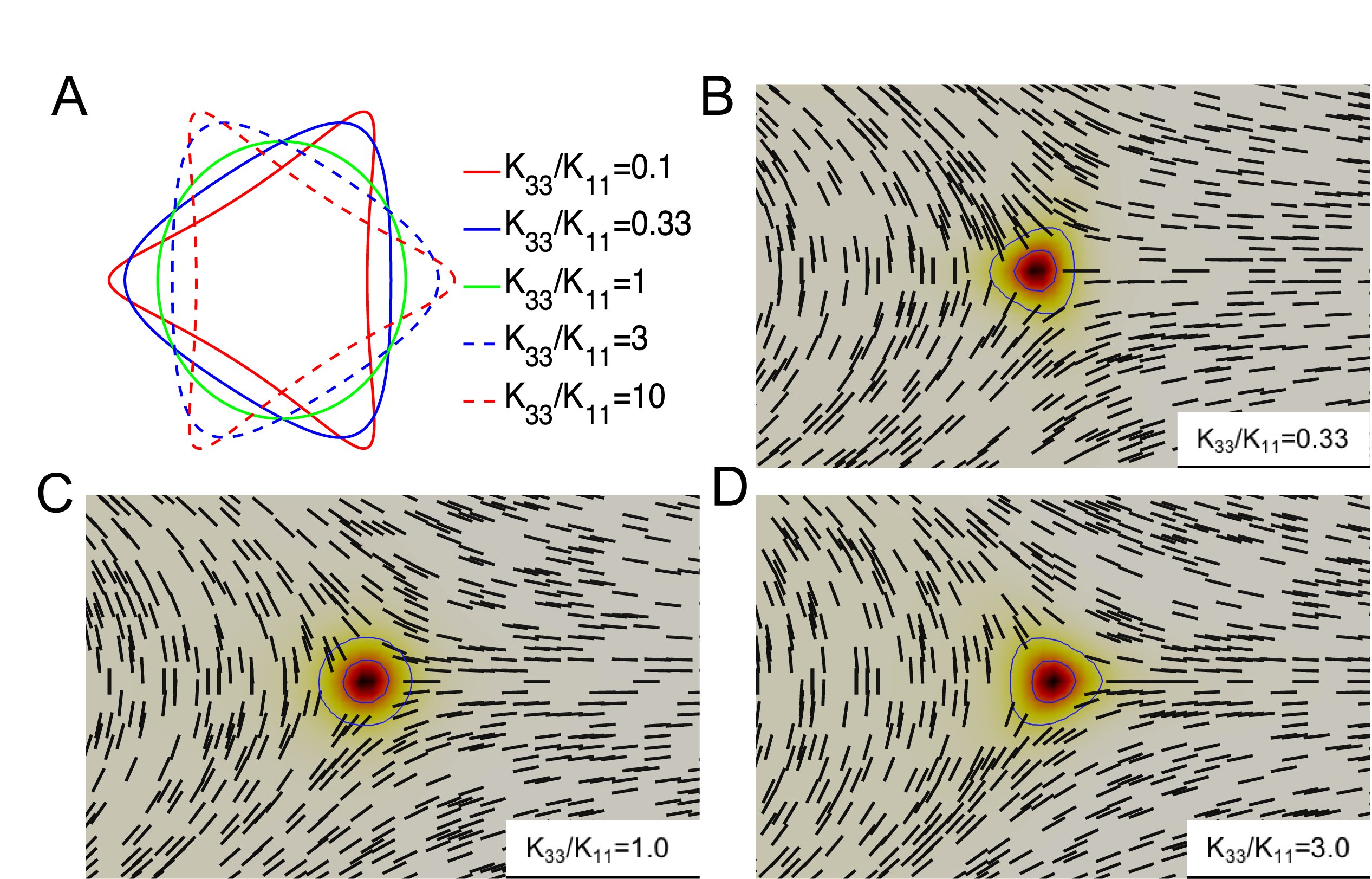}}
%\centering
%\includegraphics[width=0.85\linewidth]{s1.eps}
\caption{\textbf{Core shape of topological defects}. {\bf A}. Defect core shape calculated from the minimization of Frank-Oseen elastic energy. {\bf B}-{\bf D}. {\bf Q}-tensor simulation results of the $-1/2$ defects with various $K_{33}/K_{11}$. Iso-surfaces of the scalar order parameter are shown in black curves surrounding the defect core.}
\label{s1}
\end{figure}

Elastic constant ratio $\kappa$ also alters the core shape of $-1/2$ defects. As shown in Fig.~\ref{s1}, the defect cores adopt different shapes at different $\kappa $, despite that the director fields look similar. When $K_{33}=K_{11}$, the core periphery is a circle. Whereas when $K_{33}\neq K_{11}$, the core exhibits triangular-like shape. The orientation of such ``triangle'' depends on $\kappa $. The underlying reason is again related to the splay/bend distribution near $-1/2$ defect. It is not surprising that the defect core is circular when splay and bend are symmetric. If splay and bend are unequal, the three-fold symmetry of the splay/bend energy leads to the triangular-like core shape. When $K_{33}<K_{11}$, bend is cheap and thus the iso-surface of elastic energy advances along the high bend region (blue), resulting in the triangle's orientation pointing to the left. When $K_{33}>K_{11}$, splay is cheap and therefore the iso-surface of elastic energy advances along the high splay region (red), leading to the triangle's orientation pointing to the right.

\subsubsection*{$+1/2$ defect shape for long actin filaments}
\begin{figure}%[b]
 \centerline{\includegraphics[width=0.7\textwidth]{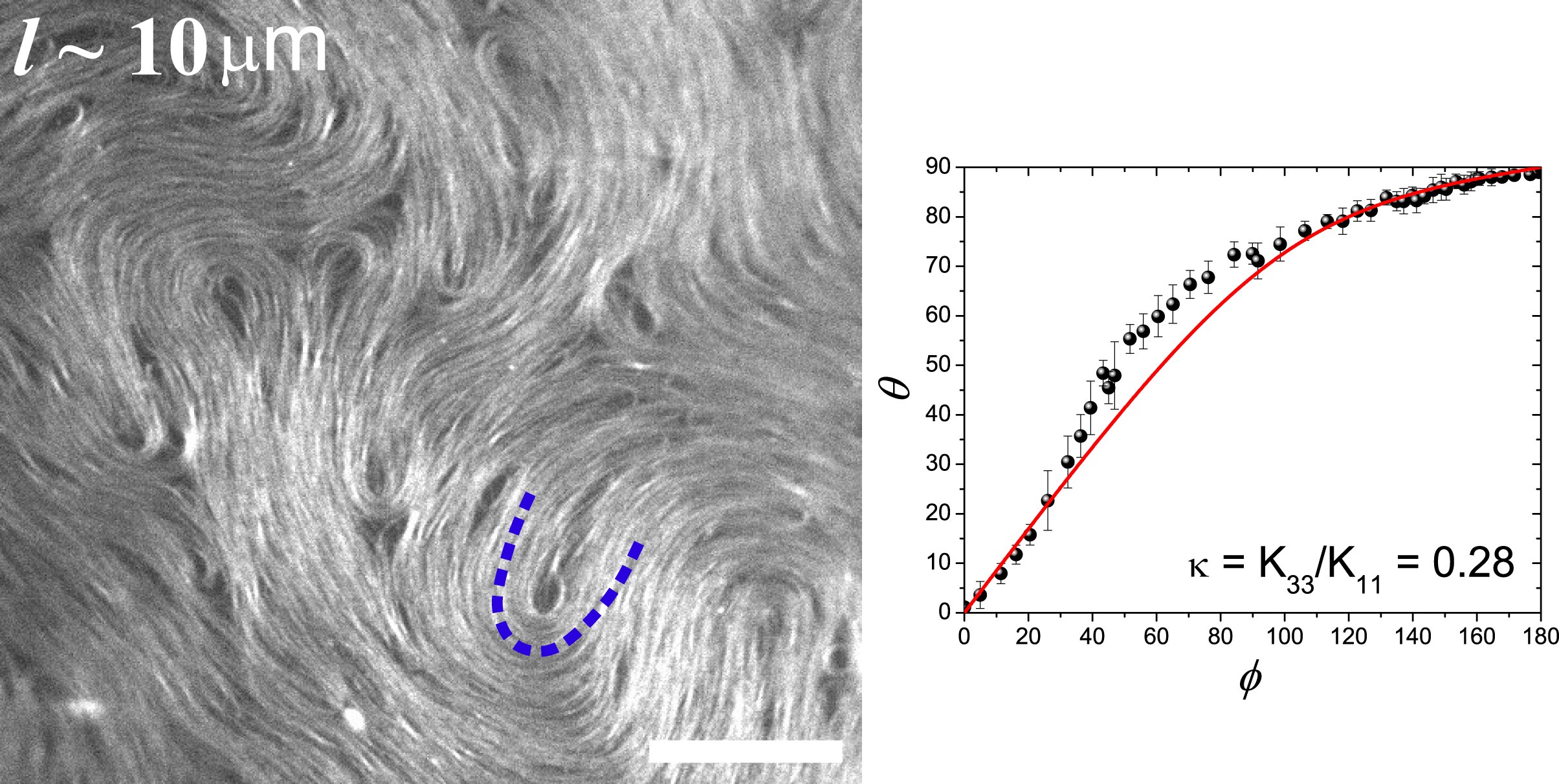}}
%\centering
%\includegraphics[scale=0.7]{LongFilament.eps}
\caption{\textbf{+1/2 defect shape for long actin filaments}. Fluorescent image of LC formed by long actin filaments. A typical +1/2 defect is highlighted by a blue dashed line and the graph shows the corresponding $\theta(\phi)$ plot for that defect. (Scale bar is 20 $\mu$m) }
\label{longactin}
\end{figure}

Here we extract $\kappa$ for the case when the actin filament length is comparable to its persistence length. We perform an experiment without adding capping protein so that the filaments grow to a length $\sim$ 10 $\mu m$ and analyze the shape of a typical +1/2 defect. The results in Fig. \ref{longactin} show that $\kappa$ = 0.28, smaller than $\kappa$ = 2.13 for $l$ = 2 $\mu$m.

\begin{figure}%[t]
 \centerline{\includegraphics[width=0.55\textwidth]{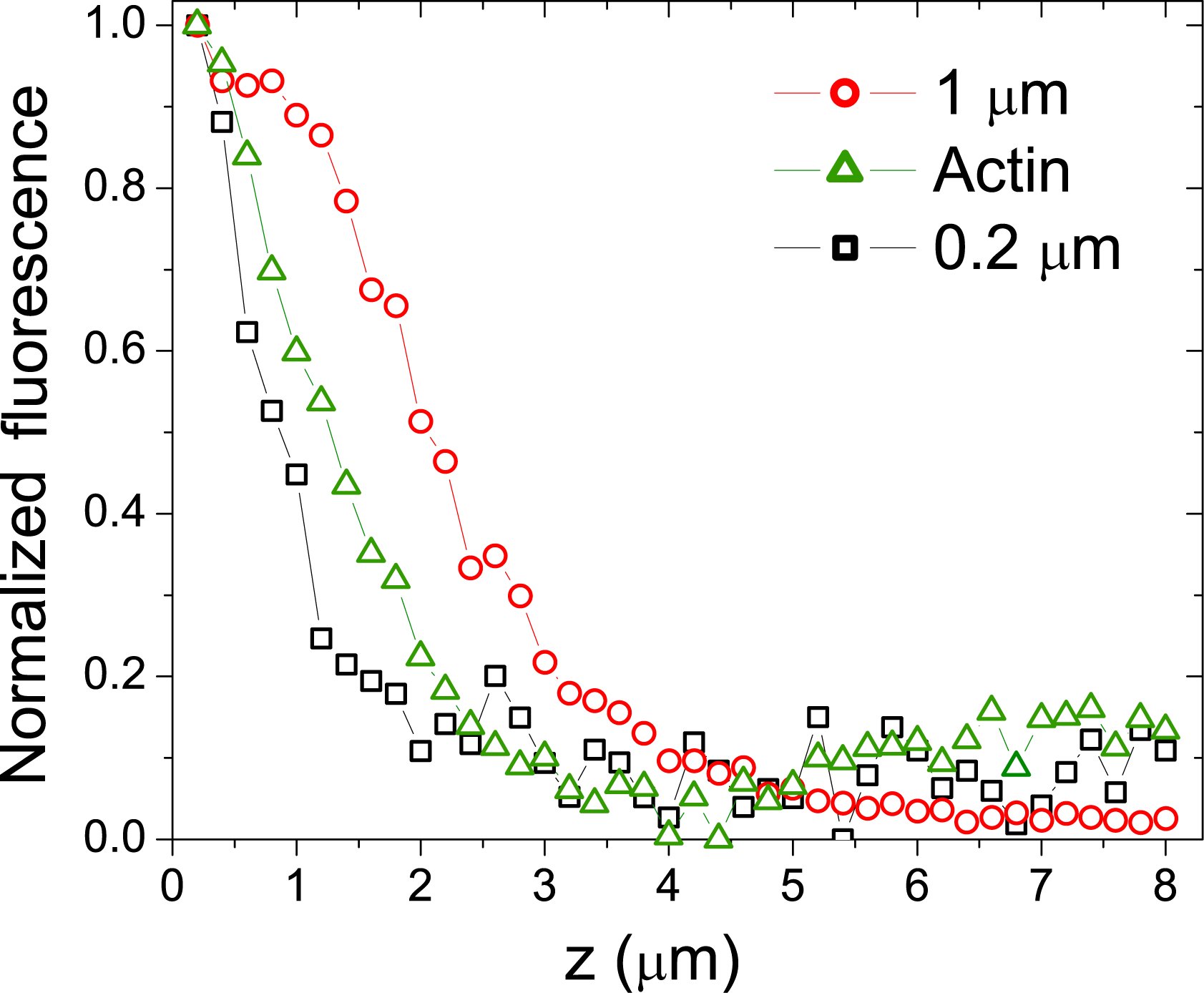}}
%\centering
%\includegraphics[scale=0.7]{s4.eps}
\caption{\label{thickness} \textbf{Actin film thickness estimate}. Green triangles show the normalized fluorescence intensity vs. distance from the oil-water interface (z) as acquired from a confocal z-stack. Comparison with the beads of diameter 1 $\mu$m and 0.2 $\mu$m stuck on a glass surface (red circles and black squares respectively) suggests that the actin film is approximately 300 nm thick.}
\label{thickness}
\end{figure}

\subsubsection*{Simple Theory on the elasticities of microtubule doped actin LC}
 \begin{figure}
  \centerline{\includegraphics[width=0.5\textwidth]{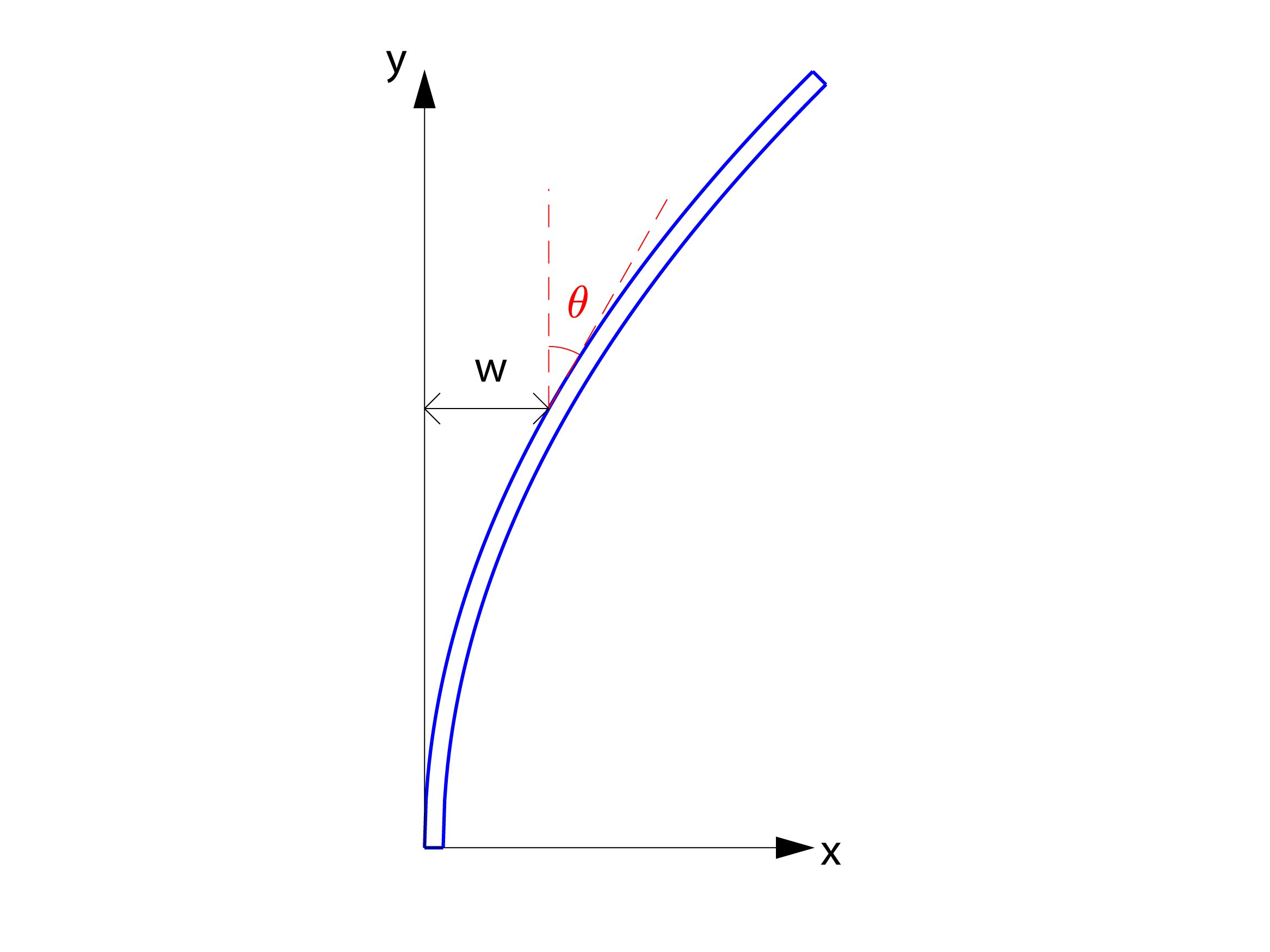}}
 \caption{\label{beam} \textbf{Elastic Beam}. The schematic of the elastic beam theory in which $w$ and $\theta$ are defined.}
 \end{figure}
The total elastic energy of the microtubule-actin composite consists of the Frank elastic energy of the nematic LC and the elastic energy of the microtubules. Elastic beam theory is adopted to describe the latter energy. The energy reads
\begin{equation} \label{mt_eq1}
\begin{aligned}
F=\int_V dV \{ & \frac{1}{2}K_{11} (\nabla\cdot {\bf n})^2+\frac{1}{2}K_{22}({\bf n}\cdot \nabla \times{\bf n})^2\\
&+\frac{1}{2}K_{33}({\bf n}\times (\nabla \times{\bf n}))^2 \} +N \int_{l_0} dy \frac{1}{2} EI \left( \frac{\partial^2 w}{\partial y^2} \right)^2,
\end{aligned}
\end{equation}
where $N$ is the number of microtubule filaments, $l_0$ is their average contour length. $w$ is the displacement (see Fig.~\ref{beam} for illustration). We assume that microtubules are aligning with the local director field ${\bf n}$, thus $w(y)=\int_0^y \tan\theta(y') dy'$. Taking the derivative to the above equation, one has $\frac{\partial w}{\partial y}=\tan \theta$. At weak bending limit, $\theta \approx 0$, therefore $\frac{\partial w}{\partial y}=\tan \theta \simeq \sin \theta$. We further have $\frac{\partial^2 w}{\partial y^2}=\cos \theta \frac{\partial \theta}{\partial y}$, thus
\[
\frac{1}{2} EI \left( \frac{\partial^2 w}{\partial y^2} \right)^2 = \frac{1}{2} EI \cos^2\theta \left(   \frac{\partial \theta}{\partial y}  \right)^2 = \frac{1}{2} EI ({\bf n}\times (\nabla \times {\bf n}) )^2.
\]
The last equator is based on writing ${\bf n}$ as ${\bf n}=(\sin\theta(y), \cos\theta(y), 0)$. One has ${\bf n}\times (\nabla \times {\bf n})=(-\cos^2\theta \frac{\partial \theta}{\partial y}, \sin\theta\cos\theta\frac{\partial \theta}{\partial y},0)$.
Therefore Eq.~\ref{mt_eq1} is rewritten as
\begin{equation}\label{mt_eq2}
\begin{aligned}
F=&\int_V dV \{  \frac{1}{2}K_{11} (\nabla\cdot {\bf n})^2+\frac{1}{2}K_{22}({\bf n}\cdot \nabla \times{\bf n})^2\\
&+\frac{1}{2}(K_{33}+c l_0 E I)({\bf n}\times (\nabla \times{\bf n}))^2 \},
\end{aligned}
\end{equation}
with $c$ the number density of microtubules, i.e. $c=N/V$. The modified bend constant becomes $K_{33}'=K_{33}+cl_0EI$. For 2D system, one has to consider the film thickness $\delta_z$, $K_{33}'=K_{33}+cl_0EI/\delta_z$, in which $c=N/A$ with $A$ the film area.

\subsubsection*{Defect annihilation}
\begin{figure}%[tbhp]
 \centerline{\includegraphics[width=1\textwidth]{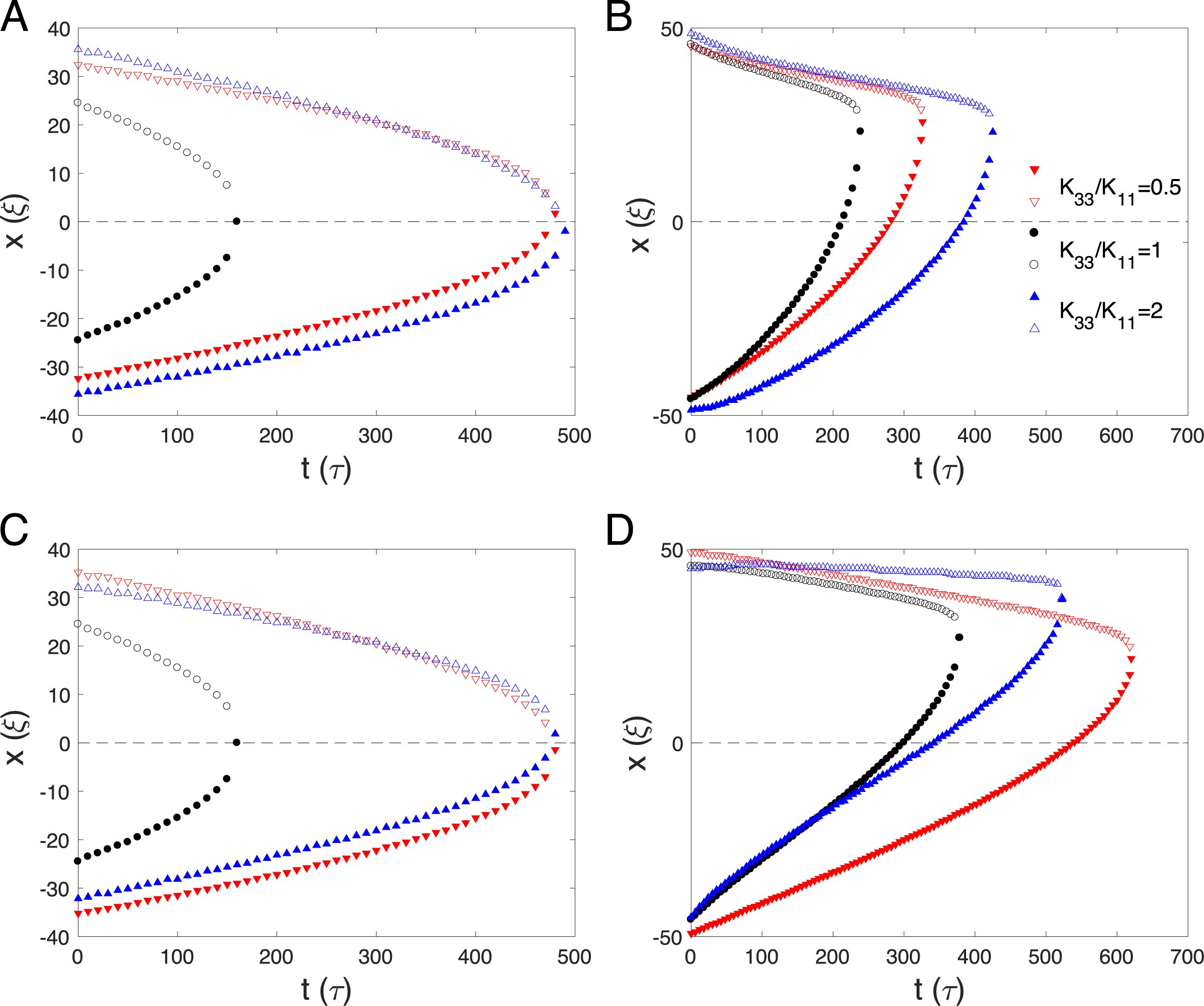}}
%\centering
%\includegraphics[width=1\linewidth]{s3.eps}
\caption{\textbf{Simulations of defect annihilation dynamics}. Defect positions are plotted against time. $0$ is where they meet if they move in the same speed. The filled symbols depict $+1/2$ defects and open symbols represent $-1/2$ defects. {\bf A}. Perpendicular case without hydrodynamic effects; {\bf B}. Perpendicular case with hydrodynamic effects; {\bf C}. Parallel case without hydrodynamic effects; {\bf D}. Parallel case with hydrodynamic effects.}
\label{s3}
\end{figure}
 In Fig.~\ref{s1}, we show the trajectories of $\pm 1/2$ defect pair during annihilation event. We consider two defect configurations with and without hydrodynamic effects. Elasticity is varied for each case. Without hydrodynamics, the relative velocity is dependent on $\kappa$ and how the defects are organized: when the defect line is perpendicular (parallel) to the far field, $+1/2$ defect moves faster (slower) than $-1/2$ defect with $K_{33}<K_{11}$. As a special case when $K_{11}=K_{33}$, the two defects move at the same speed, consistent with previous calculations\cite{toth2002hydrodynamics}. When hydrodynamic effect is included, $+1/2$ defect is more accelerated than $-1/2$ defect, again consistent with reported experimental and simulation results\cite{toth2002hydrodynamics,yanagimachi_jpsj2012}. The order of velocity anisotropy with respect to $\kappa$ is kept if hydrodynamic effect is considered: hydrodynamic flow does not change the qualitative behavior of the elasticity dependence of the defect velocity, and it only modifies the quantitative dependence. We point out that pure elastic situation is not artificial. If the 2D system is strongly confined, the long-range hydrodynamic flow can be suppressed; and the defect dynamics can be approximated at no-hydrodynamics assumption\cite{brugues_prl2008}. Therefore one should expect that at a strong confinement system, it is possible that $-1/2$ defect moves faster than $+1/2$ in an annihilation event.

\end{document}